# Cytoplasmic Viscosity is a Potential Biomarker for Metastatic Breast Cancer Cells

**Marie Dessard[1,2], Jean-Baptiste Manneville[1] and Jean-François Berret[1]***

[1]*Université Paris Cité, CNRS, Matière et systèmes complexes, 75013 Paris, France*
[2]*CNRS, INSERM, CIML, Luminy Campus, Aix-Marseille University, 13009 Marseille, France*

**Abstract**
Cellular microrheology has shown that cancer cells with high metastatic potential are softer compared to non-tumorigenic normal cells. These findings rely on measuring the apparent Young modulus of whole cells using primarily atomic force microscopy. This study aims to explore whether alternative mechanical parameters have discriminating features with regard to metastatic potential. Magnetic rotational spectroscopy (MRS) is employed in the examination of mammary epithelial cell lines: MCF-7 and MDA-MB-231, representing low and high metastatic potential, alongside normal-like MCF-10A cells. MRS utilizes active micron-sized magnetic wires in a rotating magnetic field to measure the viscosity and elastic modulus of the cytoplasm. All three cell lines display viscoelastic behavior, with cytoplasmic viscosities ranging from 10-70 Pa s and elastic moduli from 30-80 Pa. It is found that the tumorigenic MCF-7 and MDA-MB-231 cells are softer than the MCF-10A cells, with a twofold decrease in elastic modulus. To differentiate cells with low and high malignancy however, viscosity emerges as the more discriminating parameter, as MCF-7 exhibits a 5 times higher viscosity as compared to MDA-MB-231. These findings highlight the sensitivity of cytoplasmic viscosity to metastatic potential, suggesting its potential utility as a mechanical marker for malignant cancer cells.

**Keywords:** Cell mechanics - Microrheology - Breast cancer cells – Viscoelasticity - Metastatic Potential – Magnetic Rotational Spectrometry

Corresponding authors: jean-francois.berret@u-paris.fr

# I – Introduction

Cancer is responsible for 25% of deaths worldwide. Women are particularly concerned by breast cancer, which is the most common malignant disease and the second leading cause of death in this population. To stem the high mortality rate, it is essential to develop new treatments, as well as diagnostic methods for the early detection of cancerous and metastatic cells.[1-3] Malignancy transformation is known to be associated with alterations of signaling pathways regulating proliferation, metabolism, and apoptosis.[2,4,5] Indeed, when cells undergo tumorigenic transformation,[1] intermediate filaments constituting proteins keratin and vimentin are dysregulated,[6,7] the actin network becomes sparser along with fewer stress fibers,[8-11] and the relative amount of actin fibers and monomers is altered.[12] This leads to changes in cell contractility, adhesion and increased motility, facilitating cancer progression and metastatic dissemination.[13]

The tumorigenic processes are also related to changes in the mechanical properties at the tissue level.[2,12,14-16] This property has been known for a long time, to such an extent that palpation is used to detect solid tumors, because they are perceived as more rigid than the nearby tissue.[4] The perception of heightened stiffness is attributed to the dense restructuring of collagen fibers within the extracellular matrix surrounding the cancer cell aggregate.[4] However, at the cellular level, the mechanical responses remain not yet fully understood. A common idea is that the metastatic potential of cancer cells correlates with their deformability, pointing to the possibility of using specific mechanical properties as biomarkers of malignancy and cancer aggressiveness.[1,2,12] To test this hypothesis, several techniques have been developed, or adapted to cellular environments, such as atomic force microscopy (AFM),[8,14,16-26] optical stretching[13] and tweezers,[6,27,28] magnetic twisting cytometry,[29,30] micropipette aspiration,[31-34] single cell microplate rheometry[28,35] and particle-tracking microrheology.[6,9,27,36,37] In such studies, normal-like cells are compared with cancer cells of increasing invasive and metastatic potential, revealing significant differences in mechanical behavior. In this regard, research has been conducted on human breast cancer epithelial cell lines, with particular focus on the non-transformed mammary epithelial cell line MCF-10A and the tumorigenic breast cell lines, MCF-7 and MDA-MB-231, with low and high metastatic potential, respectively. Extensive microrheology studies have been carried out on these cells, setting a benchmark against which new approaches, including ours, can be evaluated. It has been showed for instance that both elastic and Young moduli are approximately twice lower in MCF-7 and MDA-MB-231 cancer cells in comparison to their healthy MCF-10A counterpart, supporting the claim that cancer cells are softer.[8,17-19,27,38,39] However, these techniques have limitations in distinguishing between cancer cells with low and high invasive and metastatic potentials, which is important information for diagnosing cancer aggressiveness effectively.

The aforementioned experiments on MCF-10, MCF-7 and MDA-MB-231 cells focused primarily on cell membrane or whole-cell deformation. However, not only cellular, but also intracellular mechanics might play a role in development of tumors.[27] Indeed, processes linked to the cytoplasmic cytoskeleton such as apico-basal polarity, cell signaling or vesicle trafficking are also disrupted in cancer.[40,41] Thus, measuring the cytoplasm mechanical properties could unravel new differences between healthy and cancer cells.

Only a limited number of studies have examined these specific intracellular properties, and their results are consistent with whole-cell experiments. Specifically, it was found that metastatic cancer cells exhibit a lower elastic modulus than in whole-cell experiments, typically a factor of 10 lower, and a higher diffusion coefficient for internalized probes.[6,9,27,36,37] Similarly, few attempts have been made to develop techniques for measuring the physical quantities used in standard rheometry[42,43] and apply them to intracellular fluids. Measuring shear stress and viscosity as a function of strain or frequency would not only make it possible to compare results with constitutive model predictions, but also to compare cells with each other in a reliable manner. Finally, one of the characteristics of current techniques, whether local or at the cellular scale, is that they are limited by design to the intermediate-to-high angular frequency range,[12,39,42] typically above $\omega$ = 0.1 rad s$^{-1}$ and do not explore the low-frequency range, *i.e.* between $10^{-3}$ and $10^{-1}$ rad s$^{-1}$. However, several studies have shown slow intracellular dynamics with characteristic times of tens or even hundreds of seconds,[44-47] and accessing the low-frequency domain would allow probing the purely viscous response of the cytoplasm.
We have recently implemented the technique of magnetic rotational spectroscopy (MRS)[48,49] and adapted it to living cells to answer the fundamental issues raised above. In MRS, magnetic micron-sized wires are used as micro-actuators and submitted to a rotating magnetic field. Analysis of wire rotation as a function of frequency or magnetic field allows us to infer values for static viscosity, elastic modulus and cytoplasmic relaxation time. Here, we have taken advantage of MRS to study the mechanical response of the well-characterized human epithelial breast cancer cells MCF-7 and MDA-MB-231, whose behavior is compared to the normal-like cell line MCF-10A. MRS data reveals significant differences in the viscoelastic properties of benign *versus* malignant cell lines. Specifically, we demonstrate that the cytoplasmic viscosity of MDA-MB-231 cells is five times lower than that of MCF-7 cells, providing a clear marker to discriminate tumorigenic cells with low and high metastatic potential.

# II – Results and discussion

## II.1 – Magnetic rotational spectroscopy: model predictions

*II.1.1 - Effect of viscosity on wire rotation*

The MRS theory has been established for anisotropic micron-sized objects with superparamagnetic properties embedded in purely viscous fluids.[50] The model was later extended by us to viscoelastic liquids and soft solids.[51] Here we outline MRS features required to analyze data on breast cancer cells. The generic behavior of a superparamagnetic wire (length $L$, diameter $D$) submitted to a rotating field can be described as follows: below a critical value $\omega_C$, the wire rotates in phase with the field in a propeller-like motion. At $\omega_C$ the wire undergoes a transition between a synchronous (S) and an asynchronous (AS) regime, the latter being characterized by back-and-forth oscillations. The critical frequency reads:

$$\omega_C = \frac{3}{8\mu_0}\frac{\Delta\chi}{\eta}\frac{B^2}{L^{*2}}$$
$$with\ L^* = L\Big/\left[D\sqrt{g(L/D)}\right] \qquad (1)$$

where $\Delta\chi = \chi^2/(2+\chi)$ denotes the anisotropy of susceptibility between parallel and perpendicular directions and , and $g(x) = ln(x) - 0.662 + 0.917x - 0.050x^2$.[52] In Eq. 1, $\mu_0$ is the vacuum permittivity, $\eta$ the fluid viscosity, $B$ the magnetic field and $L^*$ the reduced (dimensionless) length. With MRS, viscosity determination can be achieved using Eq. 1 by a single measurement of $\omega_C$ and $L^*$. Repeating the measurement with wires of different lengths, the $1/L^{*2}$-dependency can be verified, further confirming the validity of the model. This later procedure significantly increases the accuracy of the viscosity evaluation compared to a single wire measure.[53,54] In Supporting Information S1, we show that the law provided by Eq. 1 is well observed in purely viscous fluids as well as in viscoelastic liquids of Maxwell type. For the latter we used surfactant solutions that form wormlike micelles in the semi-dilute entangled regime. In addition, by selecting the surfactant concentration, the rheological parameters of the wormlike micellar solutions were adjusted so that they are of the same magnitude as those of the living cells studied in this work. Wormlike micelles can therefore be used as a reference for comparison with cells.[55] For calibration, the susceptibility parameter $\Delta\chi$ in Eq. 1 was determined by measuring the critical frequency $\omega_C$ in a liquid of known viscosity, here a $c$ = 93.6 wt. % aqueous glycerol solution at 32. 5 °C. To minimize environment differences with the active microrheology measurements on cells, the calibration was performed under strictly equivalent conditions. Details of this approach are provided in the Supporting Information S2.

*II.1.2 - Effect of elasticity on wire rotation*

For a Maxwell-type viscoelastic fluid, represented as a dashpot and spring in series, the static viscosity is expressed as the product of an elastic modulus $G$ and a relaxation time $\tau$. In such case, the synchronous/asynchronous transition still occurs and Eq. 1 is valid, provided that viscosity is rewritten $\eta = G\tau$ in the equation. The major difference between Newton and Maxwell fluid models concerns the amplitude of oscillations $\theta_B(\omega)$ in the asynchronous regime: in the viscous liquid, above $\omega_C$, $\theta_B(\omega) \sim 1/\omega$ whereas for the viscoelastic fluid, it takes a finite value in the high frequency limit, and scales with the inverse of the elastic modulus $G$:[51,53]

$$\lim_{\omega\to\infty} \theta_B(\omega) = \theta_0 = \frac{3}{4\mu_0}\frac{\Delta\chi}{G}\frac{B^2}{L^{*2}} \qquad (2)$$

Note that the relationship between the MRS viscoelastic parameters, $\theta_0 = 2\omega_C\tau$ obtained by combining Eqs. 1 and 2 is another formulation of the equality $\eta = G\tau$ known from the Maxwell model.

**II.2 – Wire structure and cell internalization**

Fig. 1a displays an optical microscopy image in phase contrast mode of as synthesized $\gamma$-$Fe_2O_3$@$PAA_{2k}$-PDADMAC magnetic wires deposited on a glass substrate. The wires are distributed in length and characterized by a median value of 8.8 ± 0.44 µm. To facilitate cellular internalization, the wires underwent a one-minute sonication, effectively reducing their size to dimensions compatible with the cellular scale. The sonicated wires were then characterized by scanning electron microscopy and energy dispersive X-ray spectroscopy mapping. The EDX image in Fig. 1b exhibits the elemental maps of iron, oxygen, carbon and silicon together, whereas the panels on the right-hand side display the elemental maps of Fe, O, C and Si elements separately. The mapping results show that Fe, C and O are present and well distributed along the wire bodies. Additional SEM and EDX results, including the proportion of each element are available in the Supporting Information S3. Size analysis of 200 sonicated wires, observed by SEM was performed to derive their length and diameter distributions (Fig. 1c), revealing median values of 4.47 µm and 0.46 µm, respectively, with a dispersity of approximately 0.6 in both dimensions (the dispersity is defined as the ratio between the standard deviation and the mean). These data are consistent with earlier determinations related to the active microrheology of living cells.[56] or pulmonary biofluids.[48,49]

Three breast cell lines with different metastatic potential are being evaluated: the normal, untransformed MCF-10A cells and the tumorigenic MCF-7 and MDA-MB-231 cells with low and high metastatic potential. Following a day of cell adhesion on the glass slide, 5×105 wires were introduced into the wells for overnight incubation, resulting in an average internalization of one wire per cell. The wires were spontaneously internalized into the cytoplasm and, as based on a previous work,[54] it is assumed that > 90% of them are in the cytosol devoid of any surrounding endosomal membrane. No differences were observed in term of microwires internalization inside the three cell lines (Supporting Information Fig. S4A). Observations made during the 2-4-hour measurement period showed that neither cell morphology nor cell cycle appeared to be affected by the presence of the wires.[57]

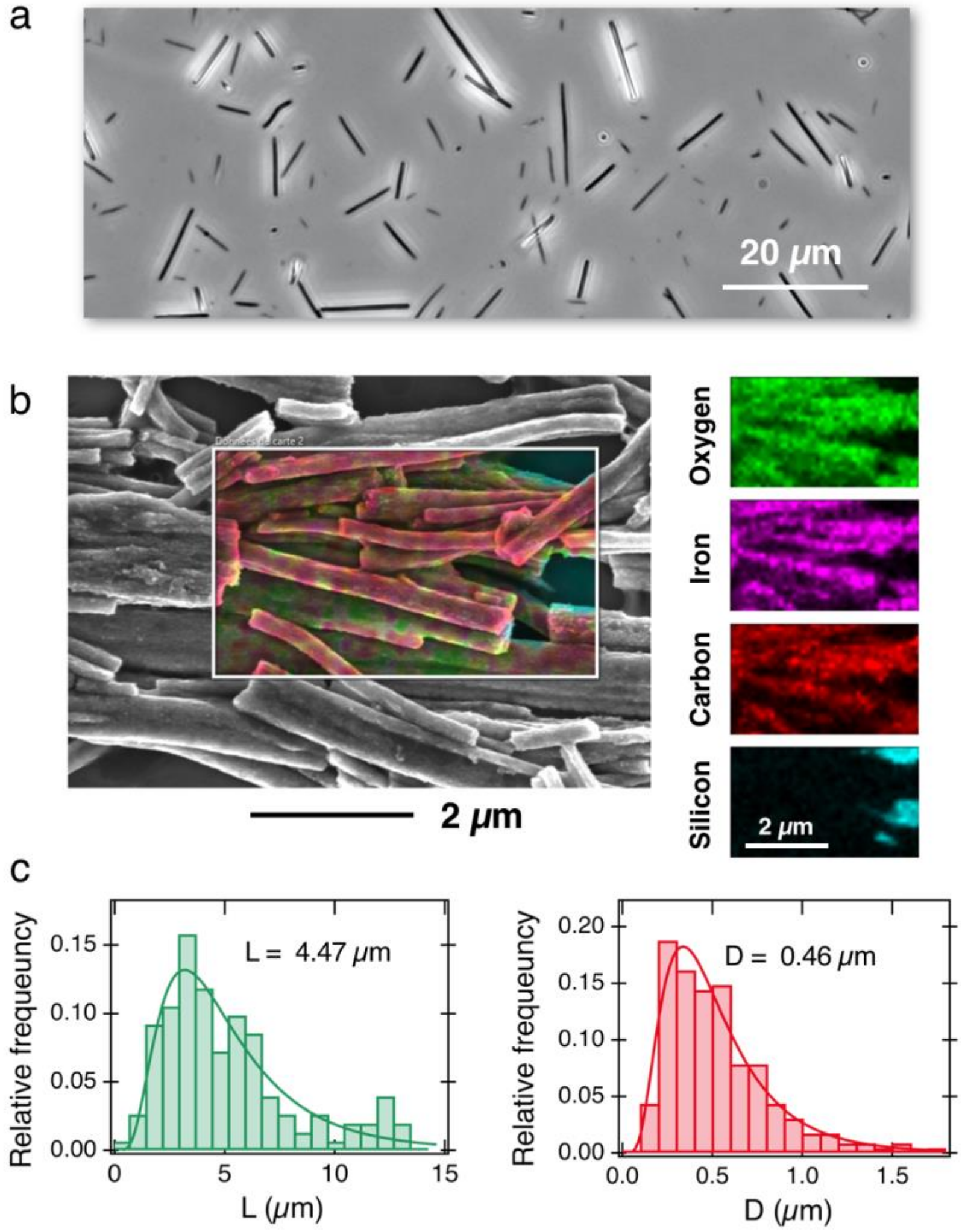

**Figure 1: a)** Phase-contrast optical microscopy image of $\gamma$-$Fe_2O_3$@$PAA_{2k}$-PDADMAC magnetic wires deposited on a glass substrate (60×). **b)** Superposition of scanning electron microscopy of magnetic wires and EDX elemental maps of iron, oxygen, carbon and silicon. For sample preparation, the wires were deposited on a silicon wafer. Right-hand panel: Individual mapping of the Fe, O, C and Si elements. **c)** Relative frequency as a function of length (**left panel**) and diameter (**right panel**) for the wires studied in this work, as determined by SEM. The continuous lines are the results of best-fit calculations using a log-normal function of median length $L$ = 4.47 µm and diameter $D$ = 0.46 µm. The dispersity of the distributions (given by the ratio between the standard deviation and the mean[58]) are 0.57 and 0.58 respectively.

### II.3 – Cytoplasm viscosity

Figs. 2a, 2b and 2c depict representative optical microscopy images of MCF-10A, MCF-7, and MDA-MB-231 cells containing magnetic wires, respectively. The wires act as internal cell actuators that can be driven remotely by an external rotating magnetic field of controlled frequency. These images are taken from sequences recorded at increasing angular frequencies from $\omega$ = 4×10$^{-3}$ rad s$^{-1}$ to 100 rad s$^{-1}$ to determine the wire rotation regime and high-frequency

oscillation behavior. Internalized wires are indicated by arrows in the figures. For MCF-10A cells, the 3.0 µm long wire seen to the right of the nucleus was monitored at frequencies $\omega$ = 0.02, 0.44 and 9.4 rad s$^{-1}$, and its motion visualized by time-lapse imaging (Supporting Information). These sequences show the succession of regimes described previously: at low frequency ($\omega$ = 0.02 rad s$^{-1}$), the wire rotates in phase with the field; at intermediate frequency ($\omega$ = 0.44 rad s$^{-1}$), its motion presents transient responses characterized by intermittent phases of rotation and oscillation; at high frequency ($\omega$ = 9.4 rad s$^{-1}$), the wire exhibits low-amplitude oscillations around a defined orientation. For this particular wire, we find a critical frequency of 0.5 rad s$^{-1}$, and an intracellular viscosity of 10.7 ± 2.7 Pa s, typical of the MCF-10A data shown below.

Figs. 2d, 2e and 2f display the critical frequency $\omega_C$ as a function of the reduced length $L^*$ for MCF-10A ($n$ = 68) , MCF-7 ($n$ = 60) and MDA-MB-231 ($n$ = 68) respectively, where $n$ indicates the number of wires investigated from at least three independent experiments. For all three cell types, a similar behavior is observed: all the wires studied do exhibit the synchronous/asynchronous transition, in agreement with the viscoelastic model prediction. The critical frequency varies according to a scaling law of the form $\omega_C(L^*) \sim 1/L^{*\alpha}$, with $\alpha$ = 7 ± 0.5 (colored straight lines). The prefactors of the $1/L^{*\alpha}$-behavior are 2400, 2100 and 11400 rad s$^{-1}$, highlighting significant differences between MDA-MB-231 cells and the other two cell lines. Interestingly, the exponent $\alpha$ does not appear to be affected by whether the cells are normal or tumorigenic breast cell lines. The $\omega_C(L^*)$-data exhibit however a deviation from the theoretical prediction derived in Eq. 1, (straight lines in grey in Figs. 2d, 2e and 2f).[48,49,53,54] In our previous work on NIH/3T3 murine fibroblasts and HeLa cervical cancer cells, discrepancies with Eq. 1 were also noted, resulting in experimental exponents $\alpha$ = -3.0 ± 0.5 and α = -6.5 ± 1.0, respectively.[56] In this first study, however, the survey sample was about 5 times smaller, with $n$ = 18 for the fibroblasts and $n$ = 10 for the cancerous cells (Supporting Information S5). The present results confirm these initial findings, with a noticeable improvement in the statistics.

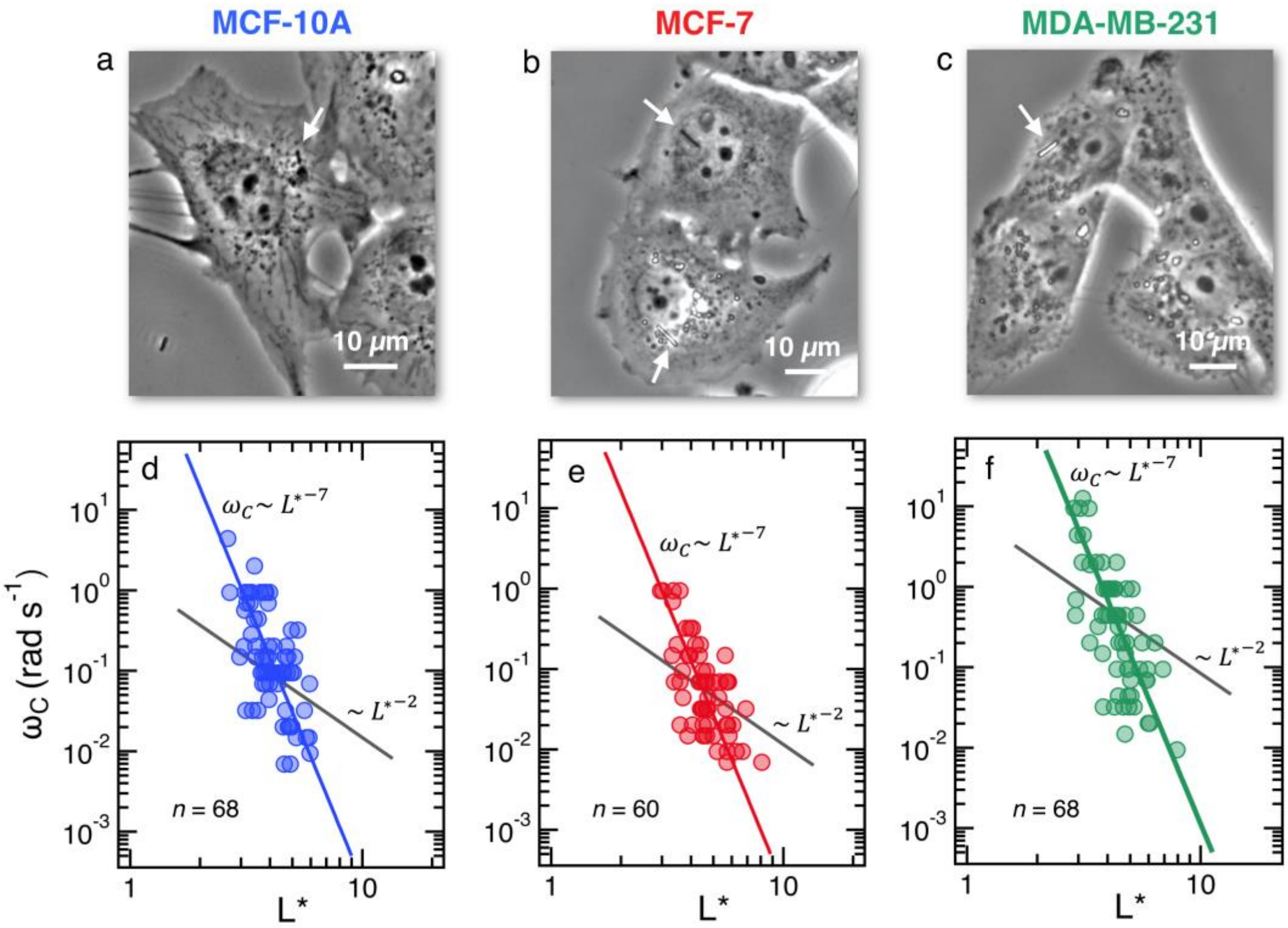

**Figure 2** : **(Upper panels**) phase-contrast optical microscopy images (×60) of **a)** MCF-10A, **b)** MCF-7 and **c)** MDA-MB-231 respectively. The arrows point to magnetic wires that have been internalized in the cells. Time-lapse animated sequences showing the different rotation regimes in MCF-10A cells can be found in Supporting Information. (Lower panels) variation of the critical frequency $\omega_C$ as a function of the reduced wire length $L^* = L/\left[D\sqrt{g(L/D)}\right]$ for **d)** MCF-10A, **e)** MCF-7 and **f)** MDA-MB-231 respectively. Straight lines with the same color as the data are least-square fits using power laws of the form $\omega_C(L^*) \sim L^{*-7}$, whereas straight lines in gray are from Eq. 1. The prefactors of the $L^{*-7}$-dependences are 2400, 2100 and 11400 rad s$^{-1}$, whereas the prefactors of the $L^{*-2}$-dependences are 1.5, 1.2 and 4.4 rad s$^{-1}$.

We now demonstrate that the $\omega_C(L^*)$-data in Fig. 2 are consistent with the assumption that intracellular viscosity depends on the wire length $L$. In accordance with the analysis outlined in Supporting Information S2, the intracellular viscosity was calculated from $\omega_C$ and $L^*$-data for each of the 196 internalized wires (Eq. S2-3). Fig. 3a shows boxplots of the viscosity of MCF-10A, MCF-7 and MDA-MB-231 cells, together with the median values and the 95% confidence intervals. The viscosities derived from this analysis lead to: $\eta_{\mathrm{MCF-10A}}$ = 36.3 ± 11.2 Pa s, $\eta_{\mathrm{MCF-7}}$ = 65.9 ± 11.4 Pa s and $\eta_{\mathrm{MDA-MB-231}}$ = 12.0 ± 5.7 Pa s. Statistical relevance using Student's t-test for unpaired samples was found significant for MDA-MB-231 cells which exhibits a 3 to 6 folds lower viscosity compared to MCF-10A and MCF-7 cells. In a second step, the viscosity data have been pooled into 4 subgroups of around 15 wires, and sorted according to their length. Figs. 3b, 3c and 3d illustrate the evolution of viscosity for such sub-groups, suggesting an effective increase of the viscosity with the probe size. To support this observation, we provide a rationale showing that intracellular viscosity does vary as a power law of the form $\eta(L) \sim L^{\beta}$ with $\beta \sim 2$ (see also Supporting Information S6). From the collected $L$- and $D$-data, it can be shown that both the diameter $D$ and the function $g(L/D)$ (Eq. 1) depend on the actual wire length $L$. The variation of the diameter with $L$ has been reported in an earlier work and was attributed to a specific feature of the wire synthesis.[48] In the range $L$ = 1-10 µm, the variation takes the form $D(L) \sim L^{0.2}$. Combining this variation with that of $g(L/D)$, one gets for the reduced length a relationship $L^*(L) \sim L^{\gamma}$ with $\gamma \sim$ ½ (Supporting Information S6). Introducing the expression for $\eta(L)$ in Eq. 1, we obtain :

$$\omega_C(L^*) \sim \frac{1}{L^{*(2+\beta/\gamma)}} \qquad (3),$$

where $2 + \beta/\gamma$ = 7 ± 0.5 from the results in Figs. 2. With the exponent coefficients found for $\gamma$, namely 0.441, 0.446 and 0.494 for MCF-10A, MCF-7 and MDA-MB-231 respectively, we finally get for the viscosity scaling law exponent $\beta_{\mathrm{MCF-10A}}$ = 2.2 ± 0.2, $\beta_{\mathrm{MCF-7}}$ = 2.2 ± 0.2, and $\beta_{\mathrm{MDA-MB-231}}$ = 2.5 ± 0.2, showing a close-to-quadratic dependence of the cell viscosity with the probe size. As shown in Fig. S6-B, as $L$ increases from 2 µm to 6 µm, the MCF-10A viscosity varies from 15 to 160 Pa s, whereas for MDA-MB-231 the variation is from 3 Pa s to 25 Pa s over the same interval. These variations are larger than the standard deviations and errors found experimentally. This behavior could be due to a confinement effect, or to the fact that the longest wires are subject to structural hindrances from organelles or plasma membrane. The previous outcome shows that a small variation of $\eta(L)$ *versus* $L$ can lead to a strong dependence of the critical frequency behavior on $L^*$. Note that for a series of biological or synthetic complex fluids studied with MRS,[48,51,53] $\eta(L)$ does not vary as a function of $L$, and $\omega_C(L^*) \sim L^{*-2}$ (Supporting Information S1). Very recently, Najafi et al studied the creep and relaxation responses of spherical beads of different sizes in the cytoplasm of living sea urchin eggs. Like us, they found that the viscoelastic properties of the cytoplasm depended significantly on probe size, a phenomenon attributed to hydrodynamic interactions between the moving object and the static cell surface.[59]

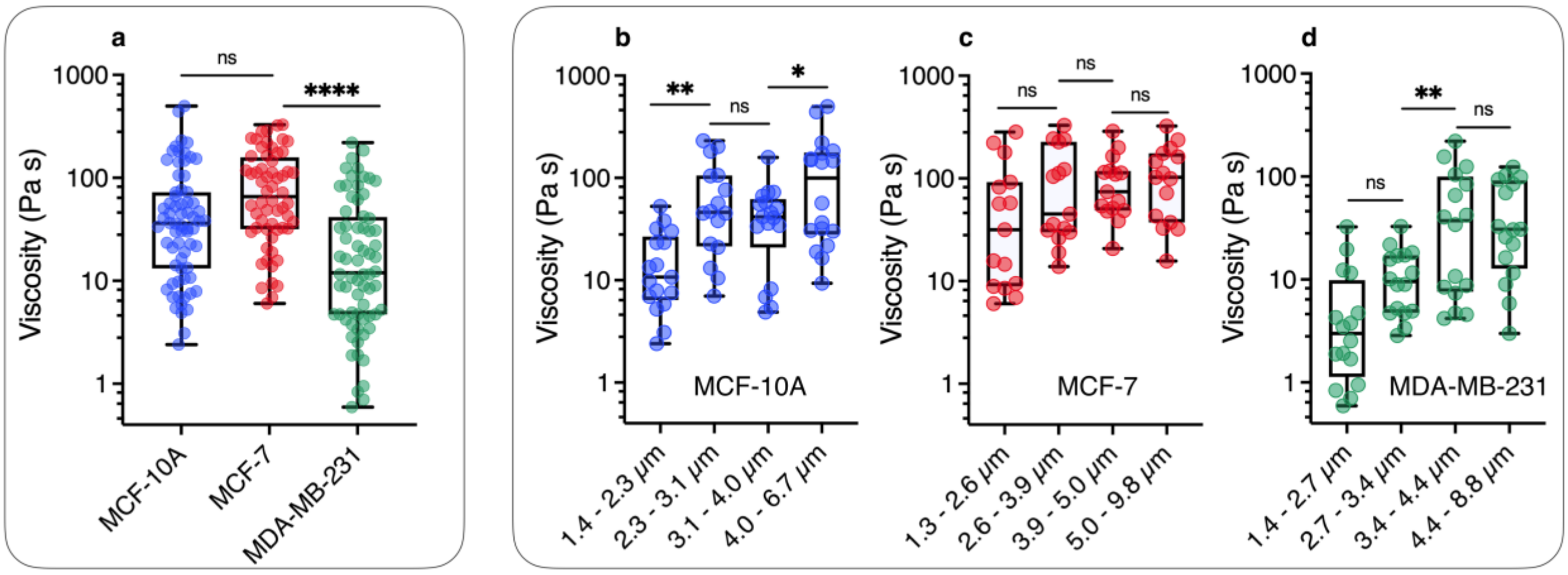

**Figure 3: a)** Static viscosity boxplots for MCF-10A ($n$ = 68), MCF-7 ($n$ = 60) and MDA-MB-231 ($n$ = 68). The median value with 95% confidence interval, and standard errors are shown on the graph. The viscosity ranges are 2.4-498 Pa s, 6.0-329 Pa s and 0.6-220 Pa s for MCF-10A, MCF-7 and MDA-MB-231 respectively. **b)** Viscosity data were distributed in 4 subgroups and sorted according to length for MCF-10A, showing a statistically significant increase between $L$ = 1.3 and $L$ = 6.7 µm. **c and d)** Similar to Fig. 3b for MCF-7 and MDA-MB-231 cell lines for length varying from $L$ = 1.3 µm and 9.8 µm. The power law of viscosity *versus* length, $\eta(L) \sim L^{\beta}$ with $\beta$ ~ 2 found for these cell lines is illustrated in Supporting Information S6.

The results of Fig. 3 have a consequence on the absolute value of intracellular viscosity: its value must be quoted for a predefined value of $L$, or at least for a narrow range of lengths. To address this issue, we focused on the length distribution of wires implemented in the MRS experiment on MCF-10A, MCF-7, and MDA-MB-231 cells (Supporting Information S4). These distributions show that a 50-60% of the wires tested have lengths between 2 and 4 µm. Fig. S4-B compares boxplots for viscosities retrieved from all available wires (circles) with that of wires with length $L$ = 3 ± 1 µm (squares). The viscosities retrieved from this second analysis led to the slightly different median values as compared to the first determination: $\eta_{\mathrm{MCF-10A}}$ = 41.6 ± 8.7 Pa s, $\eta_{\mathrm{MCF-7}}$ = 56.4 ± 16.6 Pa s and $\eta_{\mathrm{MDA-MB-231}}$ = 10.7 ± 5.4 Pa s (Table I). Comparison with the values found previously shows that the contributions of $L$ < 2 µm and $L$ > 4 µm offset each other to some extent. According to the analysis, MCF-10A and MCF-7 cells were found to have similar viscosities, while the viscosity of MDA-MB-231 cells was significantly lower, approximately 4-5 times less than the former. A key finding of this study is hence the observation that breast tumor cells with high metastatic potential demonstrate notably reduced viscosity.

| Cell lines | Viscosity $\eta$ (Pa s) | Elastic modulus $G$ (Pa) | Relaxation time $\tau$ (s) |
|---|---|---|---|
| MCF-10 | 41.6 ± 8.7 | 79.3 ± 7.3 | 0.77 ± 0.37 |
| MCF-7 | 56.4 ± 16.6 | 32.9 ± 6.0 | 3.23 ± 0.53 |
| MDA-MB-231 | 10.7 ± 5.4 | 38.6 ± 5.8 | 0.57 ± 0.20 |

**Table 1** : Median values and standard errors for static shear viscosity $\eta$, elastic modulus $G$ and relaxation time $\tau$ determined from magnetic rotational spectroscopy of MCF-10A, MCF-7 and MDA-MB-231 cells. The figures associated with these data are Fig. S4B, Fig. 5a and Fig. 6, respectively. The viscosity values are obtained for wires of length $L$ = 3 ± 1 µm.

**II.4 – Cytoplasm elasticity**

Fig. 4a illustrates the rotational motion of a 3 µm wire uptaken by a MCF-10A normal-like at frequency $\omega$ = 0.44 rad $s^{-1}$. In the upper panel, the 6 first images (t = 14-19 s) show a counterclockwise rotation, whereas the last two images reveal a more rapid clockwise return, indicating that the wire is in asynchronous mode. In the lower panel, the wire rotation angle $\theta(t)$ is displayed over the time period 0-50 s, which includes the interval corresponding to previous images. The figure also defines the oscillation amplitude $\theta_{\mathrm{B}}$, whose behavior as a function of the reduced frequency $\omega/\omega_C$ is reported in Figs. 4b, 4c and 4d for MCF-10A, MCF-7 and MDA-MB-231 cells. A dozen representative profiles are provided for each cell line, the wire lengths ranging from 2 to 8 µm. Starting at $\theta_{\mathrm{B}}$ = 1.2 ± 0.2 rad above $\omega_{\mathrm{C}}$, the angle exhibits a continuous decrease down to 0.05-0.1 rad with increasing $\omega/\omega_{\mathrm{C}}$, the decrease being stronger for MCF-10A cells. Furthermore, we observe for the three cell lines that oscillation amplitude tends towards a finite limit at high frequencies, suggesting viscoelastic behavior, in relation with Eq. 2. The figures also show the prediction for a purely viscous (Newton) fluid, for which $\theta_{\mathrm{B}}(\omega/\omega_C)$ decreases rapidly with increasing frequency, and cannot account for the intracellular data.[51,53] The data in Figs. 4 confirms the behavior already recorded for NIH/3T3 murine fibroblasts and HeLa cervical cancer cells, this time over a wider frequency range.[56] Interestingly, the wire response to the rotating field overlaps well over 3-4 decades in frequency. This independence of $\theta_{\mathrm{B}}(\omega/\omega_C)$ with wire length is found theoretically for viscous and viscoelastic model fluids.[51,53] As suggested previously,[56] it is assumed that the intracellular medium is best described as a generalized Maxwell model with a relaxation time distribution.

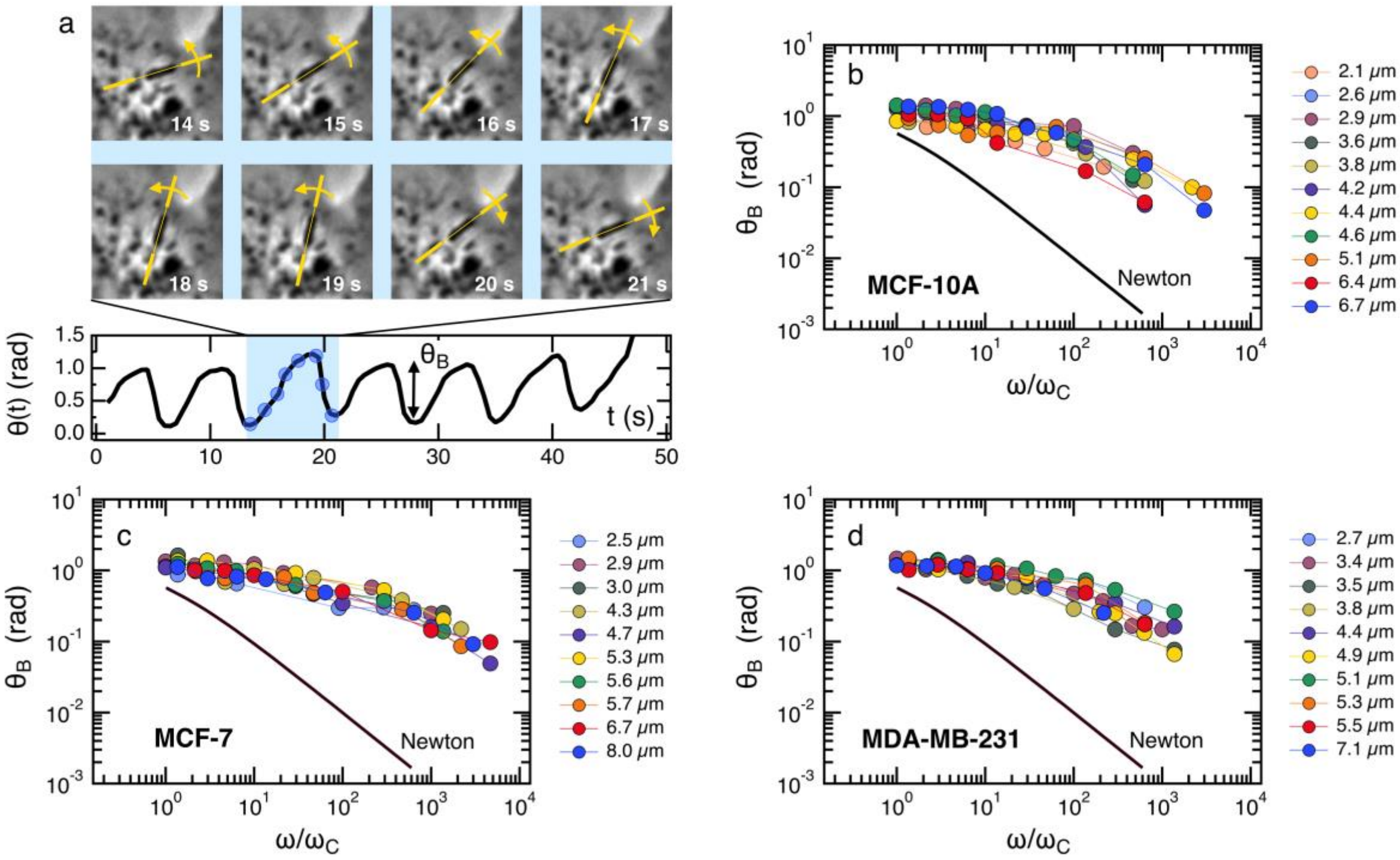

**Figure 4: a)** (**upper panel**) Optical microscopy images of a 3.0 µm magnetic wire undergoing an hindered rotation in a MCF-10A normal breast cell under the application of a magnetic field 11.5 mT and at body temperature. (**Lower panel**) Rotation angle versus time in the oscillation regime. Oscillation amplitude $\theta_{\mathrm{B}}$ in the asynchronous regime as a function of the reduced frequency $\omega/\omega_{\mathrm{C}}$ for **b)** MCF-10A, **c)** MCF-7 and **d)** MDA-MB-231 cells. For each cell, the data cover a wide range of lengths from 2 µm to 8 µm and show good superposition, suggesting a length-independent behavior. Data also show marked departure from a purely viscous fluid (continuous line labelled Newton).

From the $\theta_{\mathrm{B}}(\omega/\omega_C)$-values in the high frequency range, the elastic modulus $G$ can be derived using Eq. 2. For this, we assume that the intracellular elastic modulus is obtained from $\theta_{\mathrm{B}}$-values at $\omega/\omega_C$ = 1000, noted $\theta_B^{1000}$ for all cells, leading to $G = 3\Delta\chi B^2/4\mu_0\theta_B^{1000}L^{*2}$. This assumption slightly underestimates the actual value of the instantaneous elastic modulus, as a weak $\theta_{\mathrm{B}}$-decay is still observed above $\omega/\omega_C$ = $10^3$. Since all three cell lines show similar behavior as a function of frequency, it is assumed that the choice of $\theta_B^{1000}$ is appropriate to draw a comparison between cells. Fig. 5a shows the elastic modulus of MCF-10A ($n$ = 36), MCF-7 ($n$ = 32) and MDA-MB-231 ($n$ = 29), together with the median values and the 95% confidence intervals. The moduli derived from this analysis led to: $G_{\mathrm{MCF-10A}}$ = 79.3 ± 7.3 Pa, $G_{\mathrm{MCF-7}}$ = 32.9 ± 6.0 Pa and $G_{\mathrm{MDA-MB-231}}$ = 38.6 ± 5.8 Pa. Statistical relevance using Student's t-test for unpaired samples was found significant for MCF-7 and MDA-MB-231 compared to MCF-10A, the tumorigenic breast cells with low and high metastatic potential being 2.4 and 2.1 times softer than the normal ones (Supporting Information S7). The elastic moduli data are in good agreement with those measured by optical tweezers with 2 µm beads on MCF-10A and MDA-MB-231 cells.[27]

As for the viscosity data, the elastic modulus data series have been pooled into 4 subgroups, and sorted according to their length. Figs. 5b, 5c and 5d illustrate the evolution of elasticity for such sub-groups. The results of this analysis suggest a fairly constant elastic behavior as a function of length, in contrast to viscosity data. In a description of the intracellular environment in terms of an entangled network of filaments, the elastic modulus should vary as $1/\xi^3$, where $\xi$ denotes the network mesh size.[60] In this context, the data in Fig. 5 suggest that the cytoplasmic entities responsible for the cell mechanical behavior (cytoskeleton, organelles etc…) are homogeneous at the scale of the wires, *i.e.* between 2 and 8 µm. The elastic moduli retrieved from MRS are however 10-to-20 times lower than the apparent Young modulus obtained by AFM on the same 3 cell lines.[8,14,17-21,27,61,62] A review of these data reveals a wide spectrum of apparent Young moduli,[39] ranging from $E_{\mathrm{MCF-10A}}$ = 550 Pa[17] to 1500 Pa[20] for MCF-10A, while at the same time those for MDA-MB-231 were found between $E_{\mathrm{MDA-MB-231}}$ = 300 Pa[8] to 1000 Pa[20] (Supporting Information S7). It is interesting to note that whole-cell deformation microrheology performed with the same operating modes provide Young modulus ratios $E_{\mathrm{MCF-10A}}/E_{\mathrm{MCF-7}}$ and $E_{\mathrm{MCF-10A}}/E_{\mathrm{MDA-MB-231}}$ around 2,[8,14,17-19,27,61,62] in good agreement with the present elastic modulus ratios. This suggests that mechanical softening of tumorigenic breast cells with low and high metastatic potential is conserved at both local and cell levels.

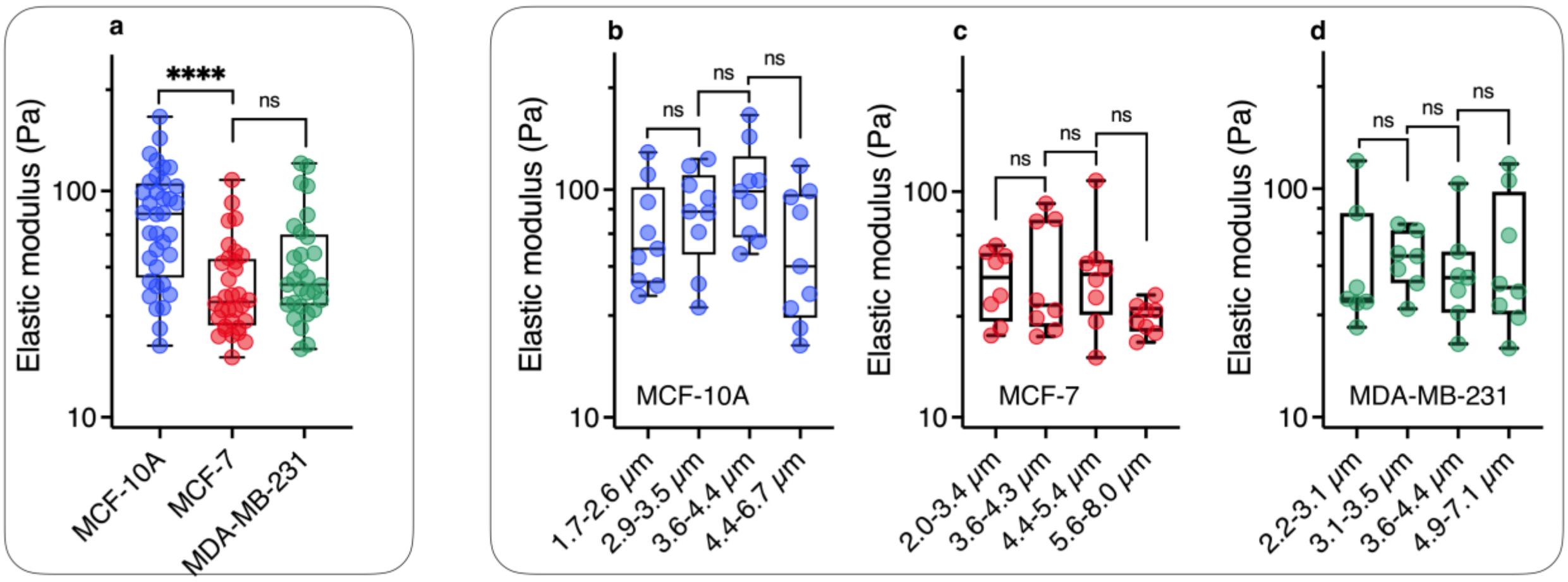

**Figure 5: a)** Scatter dot plots of measured elastic modulus $G$ for MCF-10A, MCF-7 and MDA-MB-231 cells for all wire lengths. The median value with 95% confidence interval, and standard errors are shown on the graph and in Table I. b) Elastic modulus data were distributed in 4 subgroups and sorted according to length for MCF-10A, showing a statistically non-significant variation as a function of wire length. c and d) Similar to Fig. 5b for MCF-7 and MDA-MB-231 cell lines for length comprised between $L$ = 2.0 µm and 8.0 µm.

### II.5 – Cytoplasm relaxation time

To understand how rheology can play a role in the deformation and crossing of a physical barrier by a cell, as in the case of metastatic delamination or extravasation, a third parameter, complementary to those of viscosity and elastic modulus, can be taken into account. This is the relaxation time which, in rheology, describes the temporal dynamic of stress relaxation following an applied deformation. As MRS simultaneously measures viscosity and elasticity of the volume element around the wire, it is possible to assess the relaxation time $\tau = \eta/G$ associated with this volume.[63] The individual data for each wire were used to calculate the $\eta/G$ leading to the cytoplasm relaxation time $\tau$. Fig. 6 shows boxplots for the three cell lines, which are characterized by median times 0.77 ± 0.37 s, 3.23 ± 0.53 s and 0.57 ± 0.20 s for MCF-10A, MCF-7 and MDA-MB-231 respectively (Table I). The differences are significant and indicate specific behavior, particularly between MCF-7 and MDA-MB-231 cells. If such cells were to be mechanically constrained, MCF-10 cells will deform the least, their modulus being the highest. MCF-7 and MDA-MB-231 will deform in the same way, but MDA-MB-231 will respond to the external stress $\tau_{\mathrm{MCF-7}}/\tau_{\mathrm{MDA-MB-231}}$ = 5.7 times faster than MCF-7, making them more likely to escape their environment. Measurements of entry time of human breast cells through microfluidic constrictions indicate that the MDA-MB-231 deformation kinetics is also faster than MCF-10A.[32] As can be seen, the rheological characteristics of the cytoplasm of normal breast cells and cells with increasing invasive and metastatic potential are complex. While elasticity remains an indicator of cell malignancy, viscosity and relaxation time are the most relevant marker of metastatic potential.

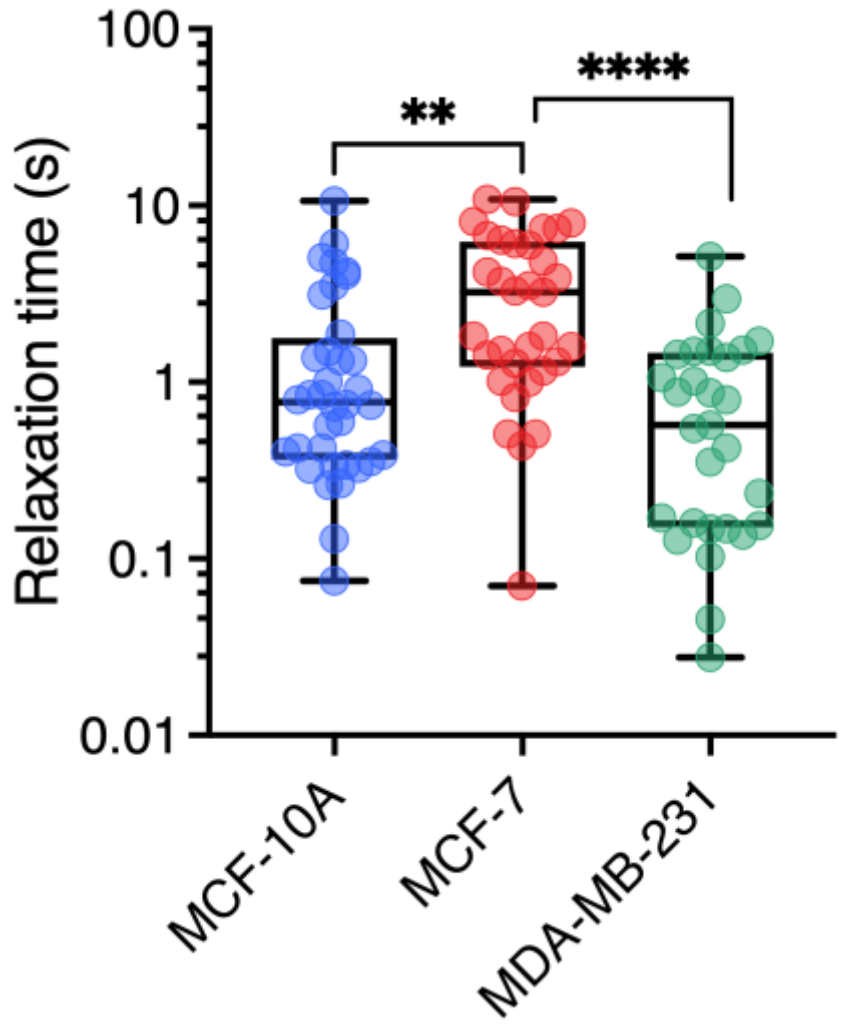

**Figure 6:** Cytoplasm relaxation time $\tau$ obtained from the ratio $\eta/G$ with $n$ = 36, 33, 29 respectively for MCF-10A, MCF-7, MDA-MB-231 respectively. Median values and standard errors are given in Table 1.

# III – Conclusion

In this work, we establish correlations between the cytoplasm mechanical properties of human breast cells MCF-10A, MCF-7 and MDA-MB-231 and their invasive and metastatic potential. Mainly studied using whole-cell deformation techniques, MCF-7 and MDA-MB-231 cell lines showed apparent Young moduli a factor of 2 lower than those of normal- MCF-10A cells. Our study aimed at investigating whether parameters such as static viscosity, elastic modulus, or mechanical relaxation time of the cytoplasm exhibit distinct patterns that could be utilized to differentiate these cells more markedly. To this end, we have used the technique of magnetic rotational spectroscopy,[53,54,64] whose protocols have been adapted to the living cell environment. Supporting Information provides detailed descriptions of these protocols. MRS probes are active magnetic wires of lengths between 1 to 10 µm submitted to a rotating magnetic field as a function of the angular frequency. MRS simultaneously measures the static viscosity and elastic modulus of the same elementary volume of cytoplasm. Predictions derived from viscous and viscoelastic model fluids, particularly with regard to wire length, have been established in previous studies, and have been applied to evaluate living cell data. MRS outcomes first confirm that the cytoplasm of MCF-10A, MCF-7 and MDA-MB-231 cells is viscoelastic, with static viscosities of around 10-70 Pa s and elastic moduli of around 30-80 Pa. For all three cell types, the wires behave similarly: the synchronous/asynchronous transition is observed for all $n$ = 196 internalized wires tested. Regarding the effect of wire length, we find that the critical frequency of the synchronous/asynchronous transition varies as $\omega_C(L^*) \sim 1/L^{*7}$ instead of the predicted $1/L^{*2}$ dependence. This result is attributed to the variation of intracellular viscosity with the probe length, for which we find a quadratic dependence, $\eta(L) \sim L^2$. To our knowledge, this is the first time such a variation has been demonstrated, and it could be due to the confining effect of the wires within the cell environment. Conversely, over the same $L$-range, we find that the elastic modulus does not depend on probe size. From the critical frequency and the amplitude of high-frequency oscillations, we then derived median values for the static viscosity and elastic modulus for the three cell lines. It is found that MCF-10A and MCF-7 have similar viscosities, around 50 Pa s, and outperform the MDA-MB-231 by a factor of 4-5. As for elastic moduli, MCF-7 and MDA-MB-231 are of the order of 35 Pa, and twice lower than those of MCF-10A, in good agreement with former optical tweezer and AFM data on the same cell lines.[11,14,18,22-25,27,36] In conclusion, our findings indicate that MCF-10A normal breast cells exhibit the highest viscosity and elasticity, while MDA-MB-231 breast tumor cells with high metastatic potential display the lowest viscosity and elasticity. Importantly, our study highlights that Young modulus is not the sole characteristic affected by the breast cancer phenotype. To differentiate cells with low and high invasiveness and malignancy, viscosity measurement proves more suitable, as it exhibits a more pronounced effect. This study hence suggests that static viscosity, instead of the elastic or Young modulus, could be used as a potential marker for invasive and metastatic cancer cells.

# IV - Materials and Methods

**IV.1 – Magnetic wire synthesis**

The wires were fabricated by electrostatic co-assembly between 13.2 nm-poly(acrylic acid) coated iron oxide nanoparticles ($\gamma$-$Fe_2O_3$@$PAA_{2k}$) and cationic poly(diallyldimethylammonium chloride) polymers (PDADMAC) of molecular weight $M_W$ = 26.8 kDa.[65] The assembly process utilizes the desalting transition. There, the two oppositely charged species, $\gamma$-$Fe_2O_3$@$PAA_{2k}$ nanoparticles and PDADMAC polymers are mixed together in the presence of an excess of salt ($NH_4Cl$ 1M), and later dialyzed against deionized water thanks to a Slide-A-Lyzer™ dialysis cassette (ThermoScientific) of cutoff 10 kDa at pH8. Unidirectional growth of the assembly is induced by a 0.3 Tesla magnetic field applied during dialysis. This technique makes it possible to manufacture magnetic wires with lengths between 1 and 100 µm and diameters between 0.2 and 2 µm.[53,56,57] With this process, micron-sized wires have inherited the characteristics of the $\gamma$-$Fe_2O_3$ nanocrystals, namely to be superparamagnetic. As shown later, this feature is essential in modeling their motion in a rotating magnetic field.[54] To be used in cells, the magnetic wires are sonicated in a sonication bath (Branson 3800, 40 kHz, 110W), thereby reducing their size to about 1-5 µm.

**IV.2 – Magnetic wire structure**

The wire structure was studied by scanning electron microscopy (SEM) and energy dispersive X-ray spectroscopy (EDX). The wire dispersion was spread on a silicon chip cleaned using a Gatan plasma cleaner to make them hydrophilic and covered by a silicon nitride amorphous film. Experiments were realized on a ZEISS Gemini SEM 360 equipped with an Oxford Instruments

Ultim Max 170 $mm^2$ detector (ITODYS Laboratoire, Paris). Samples were stuck in the sample holder using conductive double-sided adhesive tapes. All the SEM images and EDX mappings were obtained by Inlens SE detector (In Column) at 5 kV accelerating voltage. The AZtec software (Oxford Instrument) was used for the acquisition of EDX maps, point&ID analysis and line profile.

**IV.3 – Optical microscopy and environment**

For wire tracking, an Olympus IX73 inverted microscope equipped with a ×60 objective (numerical aperture 0.7) allowing bright field and phase contrast imaging was used. The data acquisition system consisted of an EXi Blue CCD camera (QImaging) working with Metamorph (Universal Imaging Inc.). Images of wires were digitized and treated with the ImageJ software and plugins (http://rsbweb.nih.gov/ij/). The rotating magnetic field was produced by a homemade device composed of two pairs of coils oriented at 90° with respect to each other, producing a 12 mT magnetic field at sample location. The signal of the two pairs of coils is phase-shifted by 90° to generate a rotating field. The current in the coils is produced by a low frequency generator coupled to a current amplifier, allowing to explore angular frequencies from $4\times10^{-3}$ rad $s^{-1}$ to 100 rad $s^{-1}$. A stream of air directed toward the measuring cell through an air inlet cover is used to thermalize the sample at 37 °C.

**IV.4 – Cell culture**

MCF-10A (ATCC- CRL-10317) is a non-tumorigenic normal-like breast human cell line. MCF-10A were grown in T25-flasks as a monolayer in Dulbecco's Modified Eagle Medium (DMEM)/F12-GlutaMAX (Gibco). The medium was supplemented with 5 vol. % horse serum, 1 vol. % penicillin/streptomycin and 1 vol. % MEGM Supplement (Lonza). MEGM is a mix of bovine pituitary extract (0.4 vol. %), recombinant human insulin-like growth factor-I (0.01 µg $mL^{-1}$), hydrocortisone (0.5 µg $mL^{-1}$) and human epidermal growth factor (3 ng $mL^{-1}$). MCF-7 (ATCC-HTB-22) is a breast tumor human epithelial-like cell line with low invasive and metastatic potential. MCF-7 were grown in DMEM supplemented with 10 vol. % fetal bovine serum (FBS) and 1 vol. % penicillin/streptomycin (PAA Laboratories GmbH). MDA-MB-231 (ATCC-HTB-26) is a breast tumor human mesenchymal-like cell line with high invasive and metastatic potential. MDA-MB-231 cells were grown in DMEM with high glucose (4.5 g $L^{-1}$), supplemented with 10 vol. % fetal bovine serum (FBS) and 1 vol. % penicillin/streptomycin. Exponentially growing cultures were maintained in T25-flasks as a monolayer in a humidified atmosphere of 5% $CO_2$ at 37°C. Cell cultures were passaged twice weekly, washed with PBS1X and detached using trypsin–EDTA (PAA Laboratories GmbH). Cells were pelleted by centrifugation at 1200 g for 5 min. The supernatant was removed and cell pellets were re-suspended in assay medium and counted using a Malassez counting chamber. Before use for an experiment with cells, wires were autoclaved at 120°C and a pressure of $2\times10^5$ Pa for 2 hours and stored at 4°C. 30mm-diameter coverslips were washed in 70% ethanol and dried. They were added in 6-well plates, washed with PBS and cell medium, and incubated at 37 °C with medium for 30 min. $5\times10^5$ cells were seeded on the coverslips. After allowing the cells to adhere for a day, $5\times10^5$ wires were added to the wells for an overnight incubation. The next day, a coverslip with cells was washed twice using warm medium and placed into the experimental PeCon device (PeCon, GmbH). A 2mm-high plastic seal was added to trap 1.6 mL medium and 24 µl HEPES buffer solution (1 M, Gibco). The device was closed using another glass coverslip and a screw.

**IV.5 – Measurement protocol**

A measurement protocol has been established for monitoring wires in living cells, taking into account the following criteria: *i)* the frequency of the rotating field must extend over a sufficiently wide interval to reveal the synchronous/asynchronous transition, *ii)* the wire number must be large to improve statistics and iii) measurements must be sufficiently rapid for the total duration of an experiment to be compatible with the type of measurements performed. We typically used $n$ = 60 wires per cell line for viscosity measurements and $n$ = 30 for elastic modulus measurements. For this, we took advantage of the fact that with the X60 objective, around 15 different cells could be simultaneously visualized, and fields of views with 5-10 wires embedded in different cells were selected and tracked. Image sequences were recorded at frequencies varying between $4 \times 10^{-3}$ rad $s^{-1}$ and 100 rad $s^{-1}$, *i.e.* over nearly 5 decades in $\omega$ frequencies, and later analyzed using ImageJ software. Measurement frequencies were $2.0\times10^{k}$, $4.4\times10^{k}$ and $9.4\times10^{k}$ rad $s^{-1}$, with k = -3 to 1. For each cell line, 3 to 5 independent experiments were carried out. The analysis of the movies provides the wire lengths $L$ and the diameters $D$, the critical frequency $\omega_C$ and the oscillation amplitude at high frequency $\theta_0$. A remarkable behavior is that the synchronous/asynchronous transition was observed over a range of frequencies noted $\Delta\omega_C$, instead of at a fixed frequency as predicted by the MRS model. This phenomenon has been attributed to the temporal variation in viscosity at the wire location.[47] In the $\Delta\omega_C$-interval, the wires have a temporal behavior of successive rotation and oscillation due to intermittent $\eta(t)$-fluctuations. For the wires showing this pattern (around 50% of the cases), $\omega_C$ has been chosen as the highest frequency where both synchronous and asynchronous regimes coexist.

**IV.6 – Statistical analysis**

All results are repeats from at least three independent experiments. More than $n$ = 30 wires were measured in all experimental conditions. Student's t-tests for unpaired samples were used to assess statistical significance and the P-values are such as: n.s, nonsignificant; *P < 0.05, **P < 0.01, and ***P < 0.001

# Supporting Information

Evidence of the $\omega_C(L^*) \sim 1/{L^*}^2$ regime in a viscous fluid and in a viscoelastic fluid (S1) – Calibration of the sonicated wire magnetic properties and determination of the cytoplasm viscosity (S2) – Complementary Scanning electron microscopy data (S3) – Sampling wires according to their length and effect on the viscosity (S4) – $\omega_C(L^*)$ versus $L^*$ data for NIH/3T3 mouse fibroblasts and HeLa cervical cancer cells, with new adjustments (S5) – Analytical derivation of the critical frequency exponent in cells (S6) – Apparent elastic and Young moduli, measured according to literature (S7)

**Movie#1** – Movie of a 3 µm magnetic wire undergoing a synchronous motion at the angular frequency of 0.02 rad $s^{-1}$ and under a magnetic field of 12 mT.
**Movie#2** – Movie of a 3 µm magnetic wire undergoing intermittent phases of rotation and oscillation at the angular frequency of 0.44 rad $s^{-1}$ and under a magnetic field of 12 mT.
**Movie#3** – Movie of a 3 µm magnetic wire undergoing asynchronous oscillations at the angular frequency of 9.4 rad $s^{-1}$ and under a magnetic field of 12 mT.
**Excel- file** – The file contains the data related to measurements made on the three cell lines MCF-10A, MCF-7, and MDA-MB-231 cells. In addition to the wire geometrical characteristics (length, diameter), values for $\omega_C$, $\theta_B$-values at $\omega/\omega_C$ = 1000 and $L^*$ are provided. Static viscosity $\eta$ and elastic modulus $G$ values are also given.

## Acknowledgments

We thank Myriam Reffay for her help with cell culture, for proofreading the manuscript and for discussions. Stimulating interactions with Mathieu Boissan and Grégory Arkowitz are also acknowledged. Sarra Derroucih, from the Itodys lab (Université Paris Cité) is also thanked for her support with SEM and EDX experiments. ANR (Agence Nationale de la Recherche) and CGI (Commissariat à l'Investissement d'Avenir) are gratefully acknowledged for their financial support of this work through Labex SEAM (Science and Engineering for Advanced Materials and devices) ANR-10-LABX-0096 et ANR-18-IDEX-0001. We acknowledge the ImagoSeine facility (Jacques Monod Institute, Paris, France), and the France BioImaging infrastructure supported by the French National Research Agency (ANR-10-INBS-04, « Investments for the future »). This research was supported in part by the Agence Nationale de la Recherche under the contracts ANR-17-CE09-0017 (AlveolusMimics), ANR-20-CE18-0022 (Stric-On), ANR-21-CE19-0058-1 (MucOnChip) and by Solvay. Financial support from ITMO Cancer of Aviesan within the framework of the 2014-2019 and 2021-2030 Cancer Control Strategy on funds administered by INSERM (grant numbers 17CP089-00 and 22CP073-00) is also acknowledged.

## TOC Image

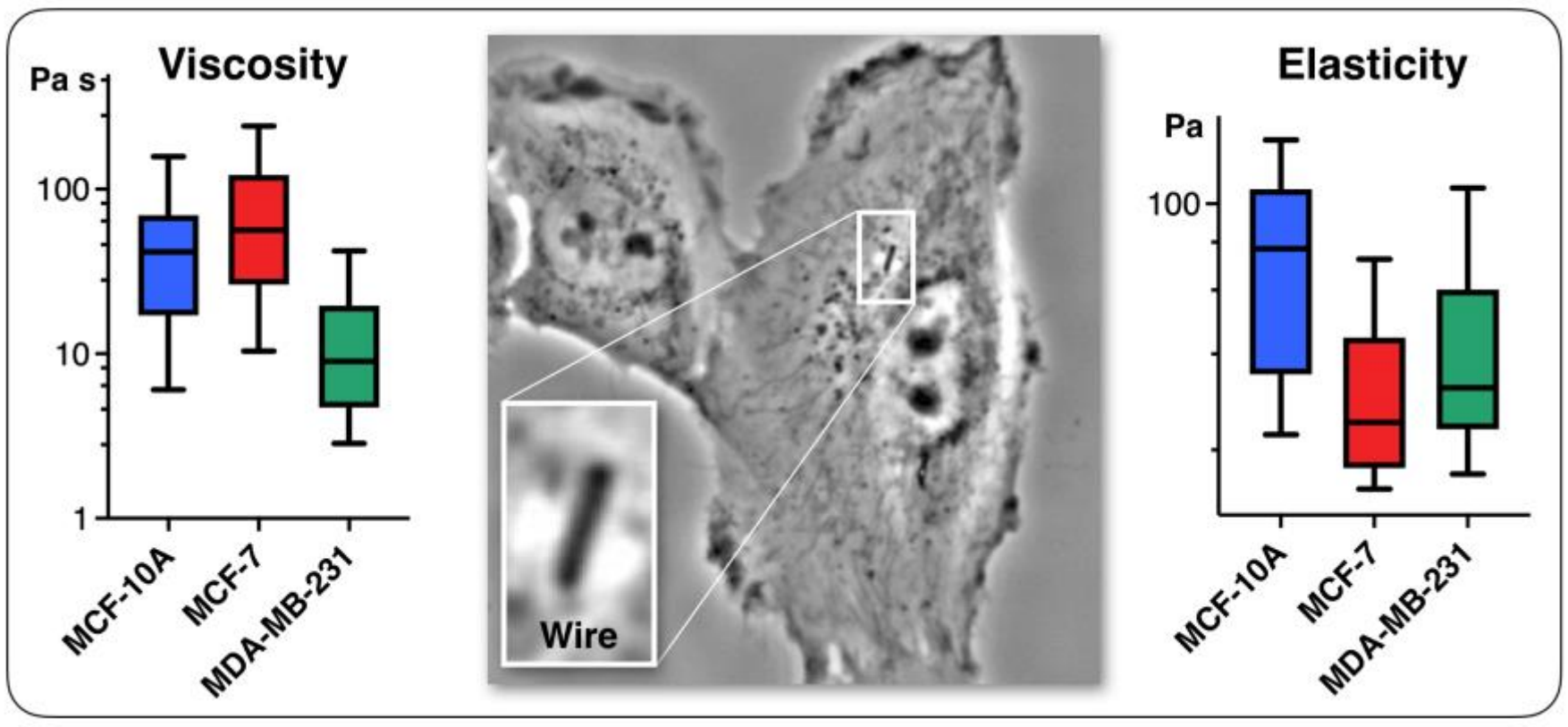